\documentclass[fleqn,useAMS,usenatbib]{mn2e}

\usepackage{graphicx}
\usepackage{amsmath}

\makeatletter

\def\lsim{\compoundrel<\over\sim}
\def\compoundrel#1\over#2{\mathpalette\compoundreL{{#1}\over{#2}}}
\def\compoundreL#1#2{\compoundREL#1#2}
\def\compoundREL#1#2\over#3{\mathrel
      {\vcenter{\hbox{$\m@th\buildrel{#1#2}\over{#1#3}$}}}}
\makeatother

\newcommand{\mbh}{M_{_{\rm{BH}}}}
\newcommand{\vw}{v_{_{\rm{w}}}}
\newcommand{\mnc}{M_{_{\rm{NC}}}}
\newcommand{\mdm}{M_{_{\rm{DM}}}}
\newcommand{\msun}{M_{\odot}}
\newcommand{\msig}{M_{\rm{\sigma}}}

\newcommand{\mbhtilde}{\widetilde{M}_{_{\rm{BH}}}}

\newcommand{\mcrittilde}{\widetilde{M}_{\rm{crit}}}

\newcommand{\Ledd}{L_{\rm{Edd}}}
\newcommand{\rsig}{r_{\rm{\sigma}}}
\newcommand{\vtilde}{\widetilde{v}}

\newcommand{\rtilde}{\widetilde{r}}

\newcommand{\mdot}{\dot{M}}
\newcommand{\edot}{\dot{E}}

\title[Black hole wind speeds and the $M$--$\sigma$ relation]{Black
  hole wind speeds and the $M$--$\sigma$ relation}
\author[R. C. McQuillin and D. E. McLaughlin]{Rachael
  C. McQuillin\thanks{E-mail: r.c.mcquillin@keele.ac.uk} and Dean
  E. McLaughlin\\ 
Astrophysics Group, Lennard Jones Laboratories, Keele University,
Keele, Staffordshire, ST5 5BG, UK} 

\begin{document}

\date{\today}

\pagerange{\pageref{firstpage}--\pageref{lastpage}} \pubyear{2012}

\maketitle 

\label{firstpage}

\begin{abstract}
We derive an $\mbh$--$\sigma$ relation between supermassive black hole
mass and stellar velocity dispersion in galaxy bulges, that results
from self-regulated, energy-conserving feedback. The relation is of
the form $\mbh \vw \propto \sigma^5$, where $\vw$ is the velocity of
the wind driven by the black hole. 
We take a sample of quiescent early-type galaxies and bulges with
measured black hole masses and velocity dispersions and use our
model to infer the wind speeds they would have had during an
active phase.   This approach, in effect, translates the scatter in
the observed  $\mbh$--$\sigma$ relation into a distribution of $\vw$.
There are some remarkable similarities between the distribution of
black hole wind speeds that we obtain and the
distributions of outflow speeds observed in local AGN, including a
comparable  median of $v_{\rm w} = 0.035c$. 
\end{abstract}


\section{Introduction}

Self-regulated feedback from accreting supermassive black holes
(SMBHs) in gaseous protogalaxies is thought to play a key role in
establishing the $\mbh$--$\sigma$ relation observed in local quiescent
galaxies, between SMBH mass and bulge-star velocity dispersion:
$\mbh \propto \sigma^x$ with $x = 4$--5
\citep{ferrarese00,gebhardt00,fandf,gultekin09,mcconnellma}.  The
accreting SMBH drives a wind, which sweeps the surrounding ambient
medium into a shell.  There is then a critical SMBH mass above which
the wind thrust pushing the shell outwards (proportional to $\mbh$) can
overcome the inward gravitational pull of the dark matter (related to
$\sigma$) and the SMBH itself.  At this critical mass, the shell may
be blown out of the galaxy, cutting off fuel to the SMBH and locking
in an $\mbh$--$\sigma$ relation
\citep{silkandrees,fabian99,king03,king05,murray}. 
Supporting this scenario are observations of strong outflows in
local active galactic nuclei (AGN), both on large scales 
\citep[e.g.,][]{sturm11} and closer to the SMBHs
\citep[e.g.,][]{pounds03,tombesi11,gofford}.  The latter
in particular
have speeds and mechanical luminosities similar to those
needed for SMBH winds to have cleared the gas from now-normal
spheroids at high redshift, when the systems were active.

The dynamics of a swept-up shell of gas depend on whether or not
the region of shocked wind material immediately behind the shell is able 
to cool.
If the shocked gas cools efficiently then the region is geometrically 
thin and the swept-up shell is pushed outwards by the ram pressure of the
wind. This momentum-driven regime is expected to be the case initially
in the case of SMBH feedback \citep{king03}, and thus many authors
have considered the $\mbh$--$\sigma$ relation that results if the
feedback is entirely momentum-driven
\citep[e.g.,][]{fabian99,king03,king05,mcq12}.
In \cite{mcq12} we considered shells moving outwards in non-isothermal,
spherical dark matter haloes that have peaked circular speed
curves.  We showed that the critical SMBH mass above which any shell
can escape tends to the limiting value (for haloes much more massive
than the SMBH, independent of any further details of the dark matter
density profile) 
\begin{equation}
M_{\rm{crit}} = \frac{f_0 \kappa}{\pi G^2} \frac{V_{\rm c,pk}^4}{4}
  \simeq 1.14 \times 10^8  \msun \left( \frac{f_0}{0.2}\right) 
  \left(\frac{V_{\rm c,pk}}{200\,\rm{km\,s}^{-1}}\right)^4 ~ .
\label{eq:msig}
\end{equation}
Here, $V_{\rm c,pk}$ is the peak value of the circular speed in the
dark matter halo; $\kappa$ is the
electron scattering opacity; and $f_0$ is a spatially constant
gas-to-dark matter mass
fraction.  The peak circular speed defines a natural
``characteristic'' velocity dispersion for a non-isothermal galaxy:
$\sigma_0 \equiv V_{\rm c,pk}/\sqrt{2}$.  Equation (\ref{eq:msig})
then implies an $\mbh$--$\sigma$ relation, which has a slope and
an intercept that are near the observed values (see
Figure 2 below). 

If the shocked gas cannot cool then the region behind the shell
is geometrically thick and hot.  The outflow is energy-driven and
the shell is pushed outwards by the thermal pressure of the shocked 
material.  In
the context of SMBH feedback, an initially momentum-driven shell is
expected to transition to energy-driven, probably quite early on when
the shell is still at relatively small galactocentric radius
(\S\ref{sec:edriven} below; cf.~\citealt{king03}, \citealt{zubovas}).
Thus, in this paper we investigate the implications of \textit{purely}
energy-driven feedback for the $\mbh$--$\sigma$ relation.

In \S\ref{sec:edriven} we derive the large-radius coasting speed,
$v_\infty$, of an energy-conserving shell in a dark-matter halo
modelled as a singular isothermal sphere with 
velocity dispersion $\sigma_0$.  We find that for the shell to coast
at the escape speed of a truncated isothermal halo (i.e.,
$v_\infty = 2\sigma_0$) requires 
\begin{equation}
\left( \frac{\mbh}{10^8 \msun} \right)
\left( \frac{v_{\rm{w}}}{c} \right) ~\simeq~
6.68 \times 10^{-2} ~ \left(\frac{f_0}{0.2}\right)\,
\left(\frac{\sigma_0}{200\,\rm{km\,s}^{-1}} \right)^5 ~,
\label{eq:intro_result}
\end{equation}
where $\mbh$ is the (fixed) SMBH mass driving a wind of speed
$v_{\rm w}$. This $\mbh$--$\sigma$ relation differs from that
resulting from momentum-driven outflows (equation [\ref{eq:msig}]),
both in the power on $\sigma_0$ and in the explicit dependence on
$\vw$.

In \S \ref{sec:gultekin} we apply our escape condition for
energy-conserving feedback to the $\mbh$--$\sigma$ relation defined
observationally by a standard sample of low-redshift, quiescent
early-type galaxies and bulges \citep{gultekin09}.   
We use equation (\ref{eq:intro_result}) to infer the black hole wind
speeds that would have had to occur during the main epoch of galaxy
and SMBH formation, if this simple model is to account for the
individual $\mbh$ and $\sigma$ values for each galaxy or bulge in 
the \citeauthor{gultekin09} sample.
This gives
a distribution of $\vw /c$ for these galaxies in the past. 
In \S \ref{sec:vwindc_dist} we compare this distribution directly to
the distributions of $\vw /c$ observed for fast outflows in different
samples of local AGN \citep{tombesi11,gofford}.  Our main result is a
remarkable similarity between these distributions.  In particular, the
median SMBH wind speed we infer for the normal galaxies
of \citeauthor{gultekin09} is
$\vw = 0.035c$, while the median of the outflow speeds in low-redshift
AGN is $\vw=0.1c$ according to \citeauthor{tombesi11} or
$\vw=0.056c$ according to \citeauthor{gofford}

\section[]{Energy-driven outflows}
\label{sec:edriven}

In the self-regulated feedback scenario the black hole wind sweeps up
a shell of ambient gas as it moves outwards. This gives rise to two
shock fronts, one propagating forwards into the ambient medium and one
propagating back into the wind material. The resulting shock pattern
has a four-zone structure: 1) the freely flowing wind; 2) the shocked
wind region lying between the wind shock and the contact surface that
separates material originally in the wind from material originating in
the ambient medium; 3) the shocked ambient medium, lying between the
contact surface and the ambient shock, also containing the original
swept-up shell that gave rise to the shock fronts; and 4) the
undisturbed ambient medium.

In detail, the dynamics of the swept-up shell depend on three
timescales: the flow time of the shell,
$t_{\rm flow} = r_{\rm s}/v_{\rm s}$, where $r_{\rm s}$ 
is the radius of the shell and $v_{\rm s}$ is the shell velocity; 
the dynamical time of the wind, $t_{\rm dyn} = r_{\rm sw}/\vw$,
where $r_{\rm sw}$ is the radius of the wind shock and $\vw$ is the
wind velocity; and the cooling time of the shocked wind,
$t_{\rm cool}$ \citep{koomckee,fgq}. 

If $t_{\rm cool} \ll t_{\rm dyn}$, then the shocked wind region cools
before more energy is injected into the region from the freely flowing
wind. The material in the region is then confined to a thin shell (so
$r_{\rm sw} \sim r_{\rm s}$) and the shell is effectively driven
outwards by a transfer of momentum from the wind impacting on its
inner side, corresponding to a momentum-driven outflow.  

If, instead, $t_{\rm cool} \gg t_{\rm flow}$, then the most recently
shocked material cannot cool in the time it takes to travel across the
shocked wind region.  The region is thick and hot and drives the shell
outwards with its thermal pressure, corresponding to an energy-driven
outflow. 

In the intermediate case,
$t_{\rm dyn} \lsim t_{\rm cool} \lsim t_{\rm flow}$, the shell is in a
partially radiative phase where most of the material cools and
condenses into a thin shell but the most recently shocked material has
not cooled and occupies most of the volume of the region.  In this
regime the outflow conserves neither energy nor momentum.

For a wind from an SMBH, with cooling primarily by inverse Compton
scattering
\citep{king03}, the cooling rate is \citep[e.g.,][]{longair}
\begin{equation}
\frac{dE}{dt} ~=~ \frac{4}{3}\, \kappa\, m_{\rm p}\, c\, u_{\rm rad}\,
  \left(\frac{v_{\rm e}}{c}\right)^2 
  \left(\frac{E}{m_{\rm e} c^2} \right)^2 ~,
\end{equation}
where $v_{\rm e}$ is the velocity of a post-shock electron; $E$ is the
post-shock electron energy; $u_{\rm rad}$ is the radiation energy
density; and $\kappa$ is the electron-scattering opacity.
We take $u_{\rm rad} = L_{\rm Edd}/(4 \pi r^2 c)$, where
$L_{\rm Edd} = 4 \pi G \mbh c/ \kappa$ is the Eddington luminosity of
a black hole of mass $\mbh$, and
$E\simeq (9/16) m_{\rm p} v_{\rm w}^2$ for the electron energy.
Then, the cooling time, $t_{\rm cool}\equiv E/(dE/dt)$, is less than
the dynamical time of the wind, 
$t_{\rm dyn} \equiv r_{\rm sw}/\vw$, at radii
\begin{multline}
r_{\rm sw} ~\la~ \frac{3}{4}\, \frac{G \mbh}{c^2}\,
  \left(\frac{m_{\rm p}}{m_{\rm e}}\right)^2 \left(\frac{v_{\rm w}}{c}\right)
  \left(\frac{v_{\rm e}}{c}\right)^2  \\
  \simeq~ 0.26~ {\rm pc}~ \left(\frac{\mbh}{10^8 M_{\odot}} \right)
  \left(\frac{v_{\rm w}}{0.03 c}\right)
  \left(\frac{v_{\rm e}}{0.85 c}\right)^2 ~.
\end{multline}
When the wind shock is inside this radius, the shocked wind region is
thin, so $r_{\rm sw} \sim r_{\rm s}$ and the shell is momentum-driven.

The cooling time exceeds the flow time of the shell,
$t_{\rm flow} \equiv r_{\rm s}/v_{\rm s}$, at radii
\begin{multline}
r_{\rm s} ~\ga~ \frac{3}{4} \frac{G \mbh}{c\,v_{\rm s}}
  \left(\frac{m_{\rm p}}{m_{\rm e}}\right)^2 \left(\frac{\vw}{c}\right)^2
  \left(\frac{v_{\rm e}}{c}\right)^2 \\
  \simeq 11~ {\rm pc}~
  \left(\frac{v_{\rm s}}{200\,{\rm km\,s}^{-1}}\right)^{-1}
  \left(\frac{\mbh}{10^8 M_{\odot}} \right)
  \left(\frac{v_{\rm w}}{0.03 c}\right)^2
  \left(\frac{v_{\rm e}}{0.85 c}\right)^2 ~,
\label{eq:r_edriven}
\end{multline}
for typical shell velocities
$v_{\rm s} \sim \sigma_0 \sim 200~{\rm km~s}^{-1}$. 
This is in rough agreement with
\citeauthor{zubovas} (\citeyear{zubovas}; see their equation [6]),
although they replace $v_s$ with an estimate for the terminal velocity
of a momentum-driven shell and normalize to a higher fiducial
$\vw$ than we do (see Sections \ref{sec:gultekin} and
\ref{sec:vwindc_dist} below for more about typical wind speeds).
In any case, the radius in equation (\ref{eq:r_edriven}) is comparable
to the sphere of influence of a $10^8 \msun$ black hole in a
stellar distribution with velocity dispersion $200~{\rm km~s}^{-1}$.
As such, SMBH outflows can be energy-conserving
over much of their evolution, and accordingly we focus on
{\it purely} energy-driven feedback in what follows.

\cite{mcl06} noted that a self-regulated feedback
scenario can also be applied to nuclear star clusters in galaxy centres
to explain the $\mnc$--$\sigma$ relation observed by \cite{f06}.
In that case, cooling by atomic processes gives a shorter
cooling timescale with a strong dependence on the wind
speed (equation [9] of \citeauthor{mcl06}), 
and a slower wind speed results in a longer dynamical time.  Thus,
outflows from nuclear clusters can cool efficiently and be
momentum-driven to much larger radii than in the black hole case.

Whether momentum- or energy-driven, the equation of motion for a shell
of swept-up gas moving out into the dark-matter halo of a protogalaxy
against the inwards gravitational pull of both the SMBH and the dark
matter behind the shell can be written as \citep[see also][]{king05}
\begin{equation}
\frac{d}{dt} \left[M_{\rm{g}}(r)v(r)\right]
 \,+\, \frac{G\,M_{\rm{g}}(r)}{r^2} \left[\mbh + \mdm(r) \right]
 ~=~ 4 \pi r^2 P ~~ .
\label{eq:newton}
\end{equation}
Here $r$ is the instantaneous radius of the shell; $\mdm (r)$ is the
dark matter mass inside radius $r$; $M_{\rm{g}}(r)$ is the mass of
ambient gas initially inside radius $r$ (i.e., the mass that has been
swept up into the shell when it has radius $r$); $v(r) = dr/dt$ is the
velocity of the shell; and $P$ is the outwards pressure on the shell.

We adopt the simple description by \citet{kingandpounds} of a 
wind driven by radiation (continuum scattering) from an accreting
SMBH, such that the wind thrust is
\begin{equation}
\mdot_{\rm out}\,\vw ~=~ \tau\,\frac{\Ledd}{c} ~~ .
\label{eq:thrust}
\end{equation}
Here $\mdot_{\rm out}$ is the mass outflow rate in the wind and $\vw$ is the
wind velocity when it escapes the black hole; these are
distinct from the mass growth rate $dM_{\rm{g}}/dt$ and the expansion
speed $v$ of the shell of swept-up ambient gas that the wind
drives.
The parameter $\tau$ is the electron-scattering optical depth in the
wind, measured down to its escape radius from the black hole (thus,
$\tau \sim\!1$ in the single-scattering limit), multiplied by a
geometrical factor (which is also $\sim\!1$) allowing 
for some non-sphericity in the wind; see \citet{kingandpounds}
for more detail. In what follows, we retain $\tau$ in our
calculations, although ultimately we assume that $\tau\approx 1$.

The pressure on the right-hand side of equation (\ref{eq:newton}) is
just the wind ram pressure,
$4\pi r^2 P = \mdot_{\rm out}\,\vw \approx \Ledd/c$, for a
momentum-driven shell. This is the case we solved in \citet{mcq12} for
isothermal and non-isothermal dark-matter halo models.
For an energy-conserving shell, the driving pressure is instead
the thermal pressure of the shocked-wind region behind the shell. In
this case, $P$ in equation (\ref{eq:newton}) satisfies the energy
equation,
\begin{multline}
\frac{d}{dt} \left[ \frac{4}{3} \pi r^3 \frac{P}{\gamma -1} \right] ~=~
\\ \null\hfill
\edot - P\frac{d}{dt}\left[ \frac{4}{3} \pi r^3 \right]
  - \frac{G\,M_{\rm{g}}(r)v(r)}{r^2} \big[\mbh +\mdm(r) \big] ~ .
\label{eq:energyeq}
\end{multline}

In this equation, $\gamma$ on the left-hand side is the ratio
of specific heats. The last three terms on the right-hand side
give the rates of work done by the expanding shell (both $PdV$ work
and the work against the gravity of the SMBH and the dark matter
behind the shell; cf. \citealt{king05}). 
The first term on the right-hand side is the rate of
energy input to the shocked wind region, which is given by the kinetic
energy flux of the wind:
\begin{equation}
\edot ~=~ \frac{1}{2} \mdot_{\rm out}\vw^2
      ~=~ \tau\,\frac{\vw}{c}\,\frac{\Ledd}{2} ~~.
\label{eq:edot}
\end{equation}
Note that this differs slightly from, e.g.,
\citet{king05,king10} and \citet{king11}, where it is either stated or
implied that $\edot = \eta \Ledd/2$ with $\eta$ the radiative
efficiency of accretion onto the black hole.
These other papers make the {\it additional} assumption
that $\mdot_{\rm out}=\mdot_{\rm Edd}=\Ledd/(\eta c^2)$. In
combination with equation (\ref{eq:thrust}) above, this requires
$\vw/c=\eta \tau$; and putting this plus $\tau\equiv 1$ into equation
(\ref{eq:edot}) is what gives $\edot=\eta\Ledd/2$. However, in this
paper we do {\it not} assume that $\mdot_{\rm out}=\mdot_{\rm Edd}$,
nor that $\vw/c=\eta\tau$ necessarily; thus, $\vw/c$ remains as an
explicit parameter in our analysis.

Now we specialise to the case of a shell expanding into a dark-matter
halo modelled as a singular isothermal sphere (SIS), with the ambient
protogalactic gas tracing the dark matter exactly.
The density of an SIS is given by $\rho_{_{\rm{DM}}}(r) =
\sigma_0^2/(2\pi G r^2)$, so that the mass inside radius $r$ is  
\begin{equation}
\mdm(r) = \frac{2\sigma_0^2 r}{G} ~ ,
\label{eq:mdm_sis}
\end{equation}
and $M_{\rm{g}}(r) = f_0 \mdm (r)$ with a fiducial (cosmic)
$f_0\approx 0.2$.
As in \citet{mcq12}, we then define characteristic mass and radius
units in terms of the characteristic velocity dispersion of the halo,
$\sigma_0$:
\begin{equation*}
\msig ~\equiv~ \frac{f_0 \kappa \sigma_0^4}{\pi G^2} 
  ~\simeq~ 4.56  \times  10^8 \msun \left( \frac{f_0}{0.2} \right)
\left(\frac{\sigma_0}{200\,\rm{km\,s}^{-1}}\right)^4
\end{equation*}
and
\begin{equation*}
\rsig ~\equiv~ \frac{G \msig}{\sigma_0^2}
  ~\simeq~ 49.25 \, {\rm pc} \, \left(\frac{f_0}{0.2}\right)
 \left(\frac{\sigma_0}{200\,{\rm km\,s}^{-1}}\right)^2 ~~.
\end{equation*}
With the identification $\sigma_0 \equiv V_{\rm c,pk}/\sqrt{2}$, the
mass unit $\msig$ is just the critical SMBH mass from equation 
(\ref{eq:msig}) for the breakout of momentum-driven shells from
non-isothermal dark matter haloes with peaked circular speed curves.
In singular isothermal spheres, the critical mass required for
mometum-driven shells to coast at large radii with the escape speed
$2\sigma_0$ is $3\msig$ (\citealt{mcq12}; see also
\citealt{silknusser}).

We eliminate $P$ from equation (\ref{eq:energyeq}) using equation
(\ref{eq:newton}), then combine with the dark-matter and gas mass
profiles of an SIS from equation (\ref{eq:mdm_sis}) and the energy
input from equation (\ref{eq:edot}), together with
$\Ledd = 4\pi G \mbh c/\kappa$.  Also,
we write $d/dt = v\,d/dr$ in order to solve for the velocity fields of
shells, $v(r)$, rather than for $r(t)$ explicitly.  Then, defining
dimensionless variables
\begin{equation*}
\widetilde{M} \equiv M/\msig\,,
\qquad
\widetilde{r} \equiv r/\rsig
\qquad \text{and} \qquad
\widetilde{v} \equiv v/\sigma_0 ~,
\end{equation*}
the equation of motion for energy-driven shells in an SIS is
\begin{multline}
\frac{d^2}{d\,\rtilde\,^2}
   \left[\rtilde\,^2 \vtilde\,^2 \,(\,\rtilde\,)\right] \,+\,
\frac{3(\gamma-1)}{\rtilde}\,
   \frac{d}{d\,\rtilde}
      \left[\rtilde\,^2\vtilde\,^2 \,(\,\rtilde\,)\right]
\\ \qquad\quad
  \,+\, 12(\gamma-1)\,\frac{\mbhtilde}{\rtilde}
   \,-\, 6(\gamma-1)\,\frac{\tau\, \mbhtilde\, \vtilde_{\rm{w}}}
                           {\vtilde\,(\,\rtilde\,)}
\\ \null\hfill
   ~=~ -4(6\gamma-5) ~.
\label{eq:eqn_motion}
\end{multline}

\begin{figure*}
\begin{center}
\includegraphics[width=80mm,angle=270]{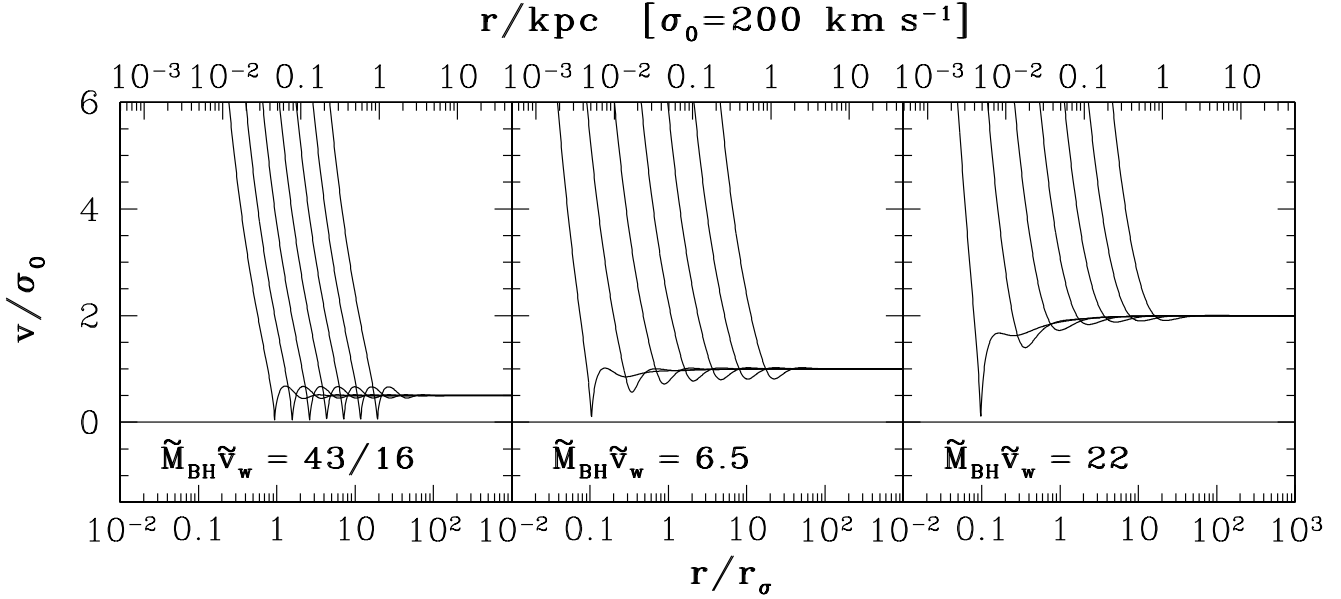}
\end{center}
\caption{Velocity fields $\vtilde$ versus $\rtilde$ that solve
  equation (\ref{eq:eqn_motion}) for energy-driven
  shells in an SIS with spatially constant gas fraction and
  $\mathbf{\mbhtilde\vtilde_{\rm{w}} = 43/16 \,(\simeq 2.7)}$, 6.5 and
  22, leading to
  large-radius coasting speeds of $v_{\infty}/\sigma_0=0.5$, 1 and 2 for
  the shells (equation [\ref{eq:sis_B}], with $\gamma=5/3$ and
  $\tau=1$). Solutions with a
  range of initial conditions are shown in each case. The
  radius unit $\rsig=49.25$~pc for a gas mass fraction $f_0=0.2$ and
  $\sigma_0 = 200\,\rm{km\,s}^{-1}$.}
\label{fig:sis}
\end{figure*}

In the limit of large radius, the term $\mbhtilde/\,\rtilde
\rightarrow 0$ in equation (\ref{eq:eqn_motion}), and the remaining
terms imply that the velocity of the shell tends to a constant: 
\begin{equation}
\vtilde ~\longrightarrow~ \vtilde_{\infty} ~ ,
 \hfill (\rtilde \gg 1)~~
\label{eq:sis_large}
\end{equation}
where $\vtilde_{\infty}$ satisfies
\begin{equation}
(3\gamma-2)\, \vtilde_{\infty}^{\,3}
  \,+\, 2(6\gamma-5)\,\vtilde_{\infty}
   ~=~ 3(\gamma-1)~ \tau\, \mbhtilde\, \vtilde_{\rm{w}} ~ .
\label{eq:sis_B}
\end{equation}
Thus, any energy-conserving shell at sufficiently large radius tends
to a coasting speed that depends on the black hole mass, the velocity
dispersion of the halo {\it and} the velocity of the black hole wind.

A natural criterion for the escape of the feedback is that it reach a
coasting speed equal to the escape speed from a truncated isothermal
sphere, $v_{\rm esc}=2\sigma_0$. Thus, we set $\vtilde_{\infty}=2$ in
equation (\ref{eq:sis_B}) and obtain a critical value for the product of
black hole mass and wind speed:
\begin{equation}
\left[\mbhtilde\, \vtilde_{\rm w}\right]_{\rm crit} ~=~
   \frac{1}{\tau}\,\frac{4(4\gamma-3)}{(\gamma-1)}
\label{eq:mvwind_gamma_1}
\end{equation}
or, with all units restored,
\begin{equation}
\big[\mbh v_{\rm{w}}\big]_{\rm crit} ~=~
\frac{1}{\tau}\, \frac{4(4\gamma-3)}{(\gamma-1)}\,
 \frac{\kappa f_0}{\pi G^2}\, \sigma_0^5 ~ .
\label{eq:mvwind_gamma}
\end{equation}
Setting $\gamma=5/3$ then gives
\begin{multline}
\left(\frac{\mbh}{10^8\msun}\right)
\left(\frac{v_{\rm{w}}}{c}\right) ~=~
\\ \null\hfill
6.68 \times 10^{-2} ~ \frac{1}{\tau}\, \left(\frac{f_0}{0.2}\right)\,
\left(\frac{\sigma_0}{200\,{\rm km\,s}^{-1}}\right)^5  ~ .
\label{eq:sis_result}
\end{multline}
This is what we will compare to the observed
$\mbh$--$\sigma$ relation in \S\ref{sec:gultekin} below. 

Equation (\ref{eq:sis_result}) shows explicitly how the escape of
energy-conserving shells from an isothermal galaxy requires
$\mbh v_{\rm{w}} \propto \sigma_0^5$ in general. If $\vw$ were
effectively the same in all galaxies (or at least uncorrelated with
SMBH mass or halo velocity dispersion), then the implication is an
observable relation $\mbh \propto \sigma^5$, as has been argued many
times \citep[e.g.,][]{silkandrees,king05}. In more detail, however,
if $v_{\rm w}$ {\it did} in fact depend on black hole mass as, say,
$v_{\rm w} \propto \mbh^y$, then equation (\ref{eq:sis_result}) would
actually imply
\begin{equation}
\mbh \propto \sigma_0^{5/(1+y)} \quad .
\label{eq:vwmbhy}
\end{equation}
That is, if $v_{\rm w}$ and $\mbh$ were correlated by even a
weak power, the logarithmic slope of the $\mbh$--$\sigma$ relation
from energy-driven outflows could differ measurably from 5. 

In the limit of small radius, equation (\ref{eq:eqn_motion}) admits
solutions of the form
\begin{multline}
\vtilde\,^2\,\rtilde\,^2 ~\longrightarrow~ 
C \,-\, 4\mbhtilde\,\rtilde
  \,-\, \frac{2(6\gamma-5)}{(3\gamma-2)}~\rtilde\,^2
  \,+\, {\cal O}(\,\rtilde\,^3) ~ ,
\\ \null \hfill
(\rtilde \ll 1)
\label{eq:sis_small}
\end{multline}
where the constant $C$ represents the square of the
shell momentum,
$[M_{\rm{g}}(r)\,v(r)]^2 \propto \vtilde\,^2\,\rtilde\,^2$,
at $\rtilde=0$. In order for equation (\ref{eq:sis_result}) to apply,
a shell moving out from $\rtilde=0$ must have an initial momentum
large enough to keep $\vtilde\,^2\,\rtilde\,^2 > 0$ and
avoid stalling before it reaches the large radii where the coasting
speed in equation (\ref{eq:sis_B}) applies.

\begin{figure*}
\begin{center}
\includegraphics[width=175mm]{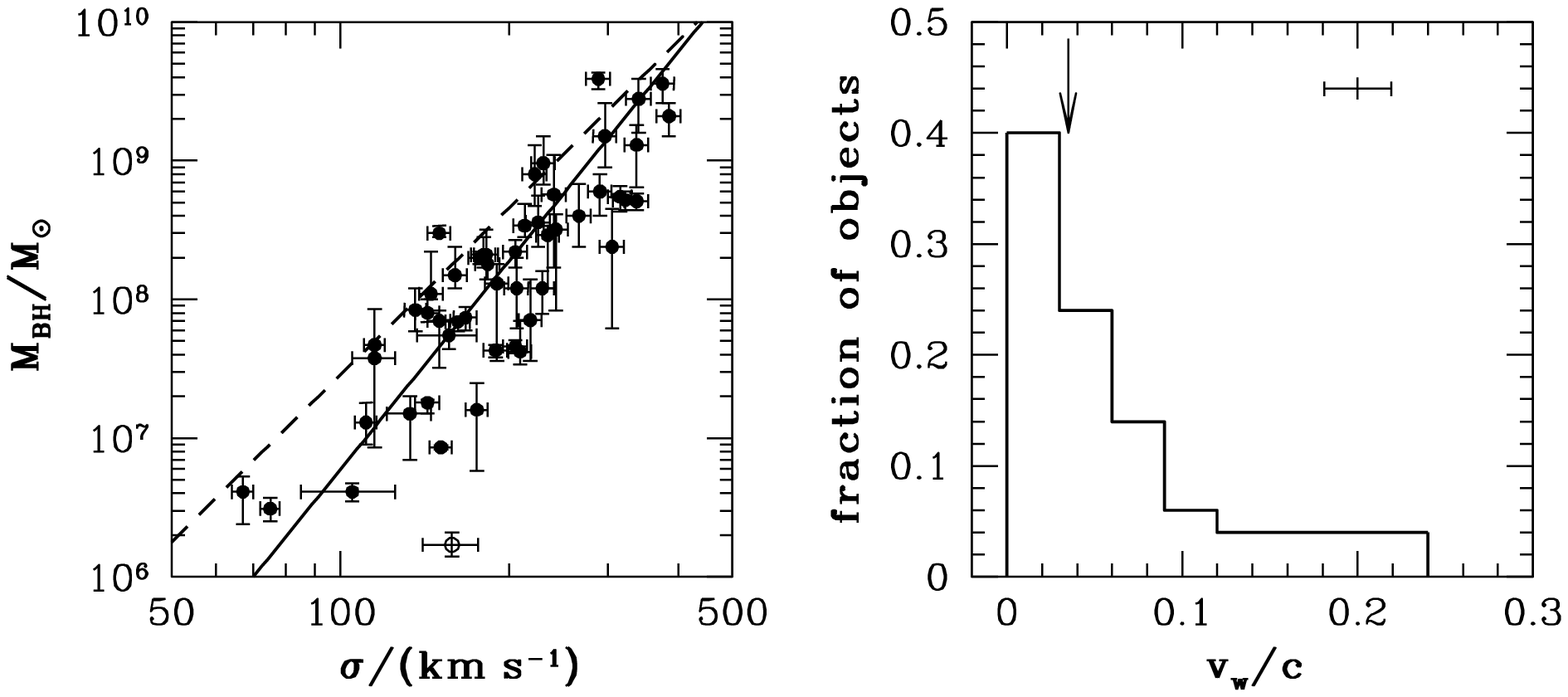}
\end{center}
\caption{\textit{Left-hand panel:} The $\mbh$--$\sigma$ relation from
  the compilation of \citet{gultekin09}.  The dashed line shows 
  $\mbh = M_{\rm crit}$ from equation (\ref{eq:msig}), the sufficient
  condition for the escape of purely momentum-driven shells from
  non-isothermal haloes \citep{mcq12}.  The solid line shows the
  condition for the escape of an energy-driven shell from an SIS from
  equation (\ref{eq:sis_result}), with $f_0 = 0.2$, $\tau =1$ and a
  typical SMBH wind speed of $\vw = 0.035c$.
  \textit{Right-hand panel:} The distribution of $v_{\rm{w}}/c$
  obtained from applying equation (\ref{eq:sis_result}) to the
  measured $\mbh$ and $\sigma$ of the \citeauthor{gultekin09} galaxies
  (excluding Circinus; see text).  The median of the distribution,
  $v_{\rm{w}} = 0.035c$, is indicated by the arrow. The errorbar
  represents the median uncertainty,
  $\Delta(v_{\rm{w}}/c)\simeq\pm 0.02$.}
\label{fig:msigma}
\end{figure*}

Figure \ref{fig:sis} shows the velocity fields,
$\vtilde\,(\,\rtilde\,)$,
that solve equation (\ref{eq:eqn_motion}) with $\gamma=5/3$, $\tau=1$
and dimensionless
$\mbhtilde\vtilde_{\rm w} = 43/16\,(\simeq\!2.7)$, 6.5 and 22. 
The different curves in each panel represent different initial
shell momenta, i.e., different values of $C$ in equation
(\ref{eq:sis_small}). We have specified a fixed wind speed in all
cases: $\vtilde_{\rm  w}=45$, which corresponds to $\vw = 0.03c$ for
$\sigma_0=200~{\rm km~s}^{-1}$. The dimensionless black hole masses
are then (again, assuming $\tau=1$) $\mbhtilde \simeq 0.06$, 0.14 and 
0.49. These are all below the critical SMBH masses for the escape of
momentum-conserving shells from either non-isothermal haloes
($\mcrittilde=1$) or an SIS ($\mcrittilde=3$).
Given any of the black hole masses
represented in Figure \ref{fig:sis}, all purely momentum-driven shells
would stall at relatively small radii and go into collapse until the
SMBH grew substantially (see Figure 1 of \citealt{mcq12}).

With $\gamma=5/3$ and $\tau=1$, equation (\ref{eq:sis_B}) gives
the final coasting speeds of the energy-driven shells illustrated in
Figure \ref{fig:sis} as $v_\infty/\sigma_0 = 0.5$, 1 and 2
(independent of initial conditions) in the three panels from left to
right. These are confirmed by our numerical solutions for the full
$\vtilde\,(\,\rtilde\,)$. In particular, all of the energy-driven
solutions in the case $\mbhtilde\vtilde_{\rm w}=22$ eventually attain
the speed for escape from a truncated SIS,
$v_\infty = 2\sigma_0 = v_{\rm esc}$. Energy-conserving feedback can
blow out of an isothermal halo if driven by a wind at speed
$\vw \sim 0.03c$ (of the order of the nuclear outflows observed in local
AGN; see below) from an SMBH significantly
less massive than that required to expel momentum-conserving shells
from isothermal or non-isothermal haloes.

\section[]{The Observed $\mbh$--$\sigma$ Relation}
\label{sec:gultekin}

The left-hand panel of Figure \ref{fig:msigma} shows $\mbh$ versus
bulge-star velocity dispersion
$\sigma$ for 51 normal (quiescent) early-type galaxies and bulges in
Table 1 of \citet{gultekin09}.\footnotemark
\footnotetext{The main outlier in Figure \ref{fig:msigma}, marked by
  an open circle, is Circinus, which is in the plane of the Milky
  Way. \citet{fandf} note that $\mbh$ in this case may be in
  error, possibly because the inclination of the maser disc used to
  find $\mbh$ is unconstrained. \citet{gultekin09} discuss this
  further.}
The dashed line on the plot traces the relation
\begin{equation}
\left(\frac{\mbh}{10^8 \msun} \right)
~=~ 4.56 ~ \left(\frac{f_0}{0.2}\right)\,
           \left(\frac{\sigma}{200\,\rm{km\,s}^{-1}}\right)^4 ~ .
\label{eq:msigma_p}
\end{equation}
This represents the SMBH mass $M_{\rm crit}$ of
equation (\ref{eq:msig}) above, which is sufficient
for the escape of any purely momentum-driven shell from any
non-isothermal dark-matter halo, {\it if} the peak circular speed in
the halo of an observed galaxy can be estimated as
$V_{\rm c,pk}=\sqrt{2}\,\sigma$ \citep{mcq12}. In a singular
isothermal sphere, for a momentum-driven shell to reach the escape
speed of $2 \sigma$ at large radii requires
$\mbh \geq 3\,M_{\rm{crit}}$---that is, SMBH masses a further 0.5~dex
above the dashed line in Figure \ref{fig:msigma}, which already
represents an upper limit to the data.
Relaxing the isothermal assumption alleviates some of this 
difficulty, and additional momentum input from bulge-star formation
triggered by the outflow could further reduce the requirement on
$\mbh$ from that in equation (\ref{eq:msigma_p}) (see, e.g.,
\citealt{silknusser}; and further discussion in \citealt{mcq12}).

By contrast, the solid line running through the data in Figure
\ref{fig:msigma} is the SMBH mass required for energy-driven shells to
escape singular isothermal spheres, from equation
(\ref{eq:sis_result}) with a fixed SMBH wind speed of $\vw/c=0.035$
(and assuming a wind optical depth $\tau=1$ and a gas-to-dark matter
mass fraction $f_0=0.2$). With $\vw/c$ set to a constant
to draw this line, it has a slope $\mbh\propto \sigma^5$, the usual
expectation for energy-conserving feedback. The numerical value of
$\vw/c$ then sets the intercept, and the value that we have applied is
in fact the median of a {\it distribution} of wind speeds that we have
estimated individually for every galaxy in \citet{gultekin09}.

These are all quiet, non-active galaxies and bulges. But if their
black hole masses were frozen in as part 
of the feedback process clearing ambient gas from the proto-spheroids,
and if this feedback was energy-driven, then equation
(\ref{eq:sis_result}) can be used to infer the SMBH wind
speeds in the past, when the galaxies were young and active. For
each point in the left-hand panel of Figure \ref{fig:msigma},
we have taken the measured values of $\mbh$ and $\sigma$ (and set
$\tau=1$, $f_0=0.2$) to solve equation (\ref{eq:sis_result}) for
$\vw/c$. The results are shown as the normalised histogram in the
right-hand panel of Figure \ref{fig:msigma}. The arrow there points to
the median speed, $\vw/c=0.035$. The minimum is $\vw/c=0.005$, and the
maximum (with Circinus excluded) is $\vw/c=0.23$.\footnotemark
\footnotetext{Applying this procedure to Circinus gives
  $\vw/c\simeq 1.2$ for that galaxy. In this case, the published SMBH
  mass estimate would have to be higher by a factor of
  $\ga\!4$ (for the same $\sigma$) or the stellar velocity dispersion
  lower by a factor $\approx\!1.3$ (for $\mbh$ fixed) to bring the inferred  
  wind speed down to $\vw/c \la 0.3$.
}
Uncertainties in the $\vw/c$ values follow from the uncertainties in
$\mbh$ and $\sigma$ tabulated by \citeauthor{gultekin09}, and the 
median errorbar, $\Delta(\vw/c)\simeq \pm0.02$, is also shown in
Figure \ref{fig:msigma}. 

\begin{figure}
\begin{center}
\includegraphics[width=83mm]{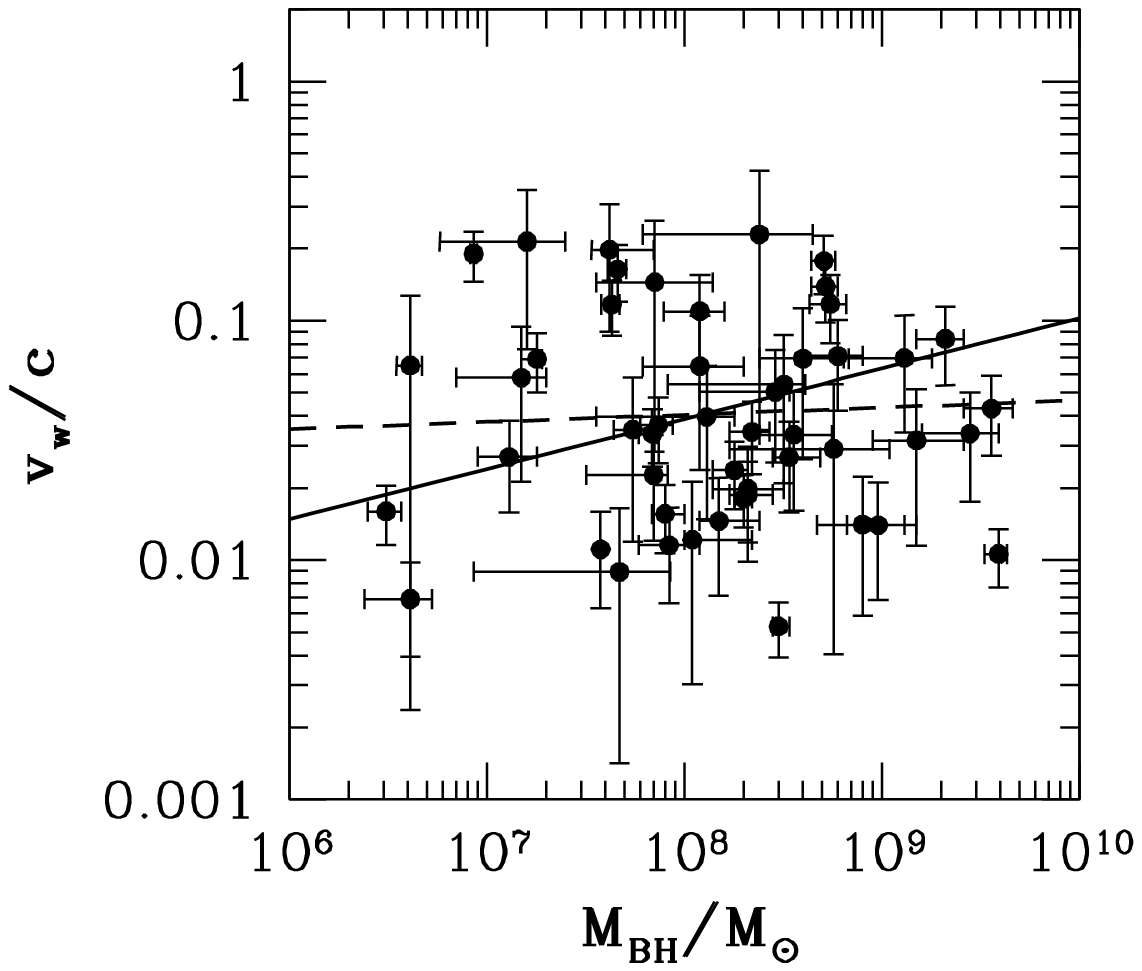}
\end{center}
\caption{
  Inferred $\vw/c$ vs.~observed $\mbh$ for the normal early-type
  galaxies and bulges in \citet{gultekin09}, with $\vw/c$ obtained 
  from equation (\ref{eq:sis_result}) for each $(\mbh,\sigma)$
  measurement. The solid line shows the correlation
  $\vw \propto \mbh^{0.2}$, which could explain the slope of
  the best-fit power-law $\mbh$--$\sigma$ relation according to 
  \citet{gultekin09}. The dashed line shows the weaker correlation
  $\vw \propto \mbh^{0.03}$, suggested by the steeper
  power-law fit to $\mbh$ versus $\sigma$ by \citet{fandf}.}
\label{fig:gultekin2}
\end{figure}

It is often reported that power-law fits to $\mbh$--$\sigma$ data
return exponents that are closer to 4 than to 5; and, as we noted in
\S\ref{sec:edriven}, even a weak correlation 
between black hole mass and wind speed could result in an
$\mbh$--$\sigma$ relation from energy-conserving feedback having a 
slope $<\!5$ (equation [\ref{eq:vwmbhy}]). Thus,
Figure \ref{fig:gultekin2} plots our inferred $\vw/c$ for the
\citeauthor{gultekin09} spheroids against their $\mbh$ values.

\begin{figure*}
\begin{center}
\includegraphics[width=150mm]{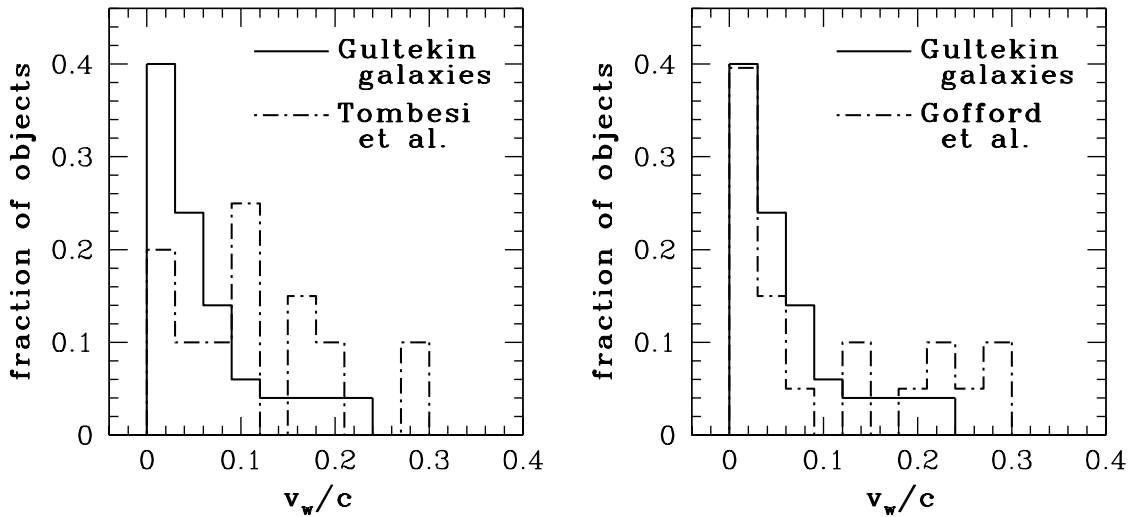}
\end{center}
\caption{
  Distribution of model (past) SMBH wind velocities for the normal
  galaxies in \citet{gultekin09} (solid lines), compared to
  the observed distributions of $\vw/c$ in local AGN, in samples
  measured by \citet{tombesi11} (dashed line in the left-hand panel)
  and by \citet{gofford} (dashed line in the right-hand panel).}
\label{fig:vwind_hist}
\end{figure*}

The fitted $\mbh$--$\sigma$ relation of
\citeauthor{fandf} (\citeyear{fandf}; their equation [20]) is
$\mbh \propto \sigma^{4.86\pm0.43}$.
Taking this power at face value, our equation (\ref{eq:vwmbhy})
implies, roughly, $\vw\propto \mbh^{0.03\pm0.09}$. This scaling is
drawn as the dashed line in Figure \ref{fig:gultekin2}. On the other
hand, \citet{gultekin09} quote $\mbh\propto \sigma^{4.12\pm0.37}$ for a
fit to the galaxies in their sample minus Circinus (see the note in
their Table 1). Putting this into
equation (\ref{eq:vwmbhy}) above implies
$\vw \propto \mbh^{0.2\pm 0.1}$, which is illustrated by the solid
line in Figure \ref{fig:gultekin2}. Either of these relations between
$\vw$ and $\mbh$ appears consistent with the data;
alternatively, neither is obviously required by the data. In fact, the
Spearman rank-correlation coefficient for these $(\vw, \mbh)$ numbers is
just $s=-0.03$, with a significance of only $\simeq\!15\%$. Any
correlation that there might be between $\vw$ and $\mbh$ is simply so
weak as to be swamped by the scatter in the $\mbh$--$\sigma$ data.
This is not really surprising but is, of course, bound up with the
well-known fact of the small {\it intrinsic} scatter in the
$\mbh$--$\sigma$ relation.

Our analysis of the $\mbh$--$\sigma$ data has essentially interpreted
the scatter in it (i.e., the spread of $\log\mbh$ at
a given $\log\sigma$) as the result of variations in SMBH wind
speeds around an average $\vw/c\simeq 0.035$. However, the histogram
in Figure \ref{fig:msigma} and the spread of the points in Figure
\ref{fig:gultekin2} have not been corrected in any way for measurement
errors in $\mbh$, which work to broaden the true, error-free
distributions. To attempt any correction is not in the scope of this
paper. But it is worth noting that the standard deviation of
our $\log(\vw/c)$ values is $\epsilon \approx 0.4$~dex,
as against an rms errorbar of $\Delta\log(\vw/c)\approx 0.1$~dex
(which is dominated by the uncertainties in $\log\mbh$).
This implies that there is indeed some real width to our $\vw/c$
distribution, which is rightly comparable to the intrinsic 
scatter in the observed $\mbh$--$\sigma$ relation
($\epsilon_0\approx 0.44$ dex, according to \citealt{gultekin09}).

In any case, the main and most robust result here is our value for the
median black hole wind speed, $\vw/c=0.035$. This not only gives a
very credible fit of a simple energy-driven feedback model to the
$\mbh$--$\sigma$ relation; it is also similar to
the typical speeds of nuclear outflows in samples of nearby,
currently active galaxies having no overlap with the
\citeauthor{gultekin09} sample of quiescent early types and bulges.

\section[]{Observed AGN outflow velocities}
\label{sec:vwindc_dist}

Highly ionised, ``ultra-fast'' outflows have been observed from the
centres of many local active
galactic nuclei since the prototypes of the phenomenon were found by
\citet{pounds03} and \citet{reeves03}. These outflows are very
massive and have high kinetic powers of the order needed, in
simple scenarios of the type discussed in this paper, for the clearing
of gaseous protogalaxies by SMBH-powered winds. As pointed
out originally by \citet{king03}, they appear to be an
observable, present-day analogue of the processes that may have worked
to establish the $\mbh$--$\sigma$ relation among now-inactive
galaxies.

Two recent studies, by \citet{tombesi11} and \citet{gofford}, give
the velocities for samples of 20 and 21 AGN outflows respectively,
with 6 sources in common. We can now compare the distributions of
these observed outflow speeds to the distribution that we inferred
in \S\ref{sec:gultekin} for SMBH wind speeds in the past, in the
normal spheroids that define the $\mbh$--$\sigma$ relation.

Figure \ref{fig:vwind_hist} shows this comparison, with the AGN
outflow velocity distribution from \citet{tombesi11} in the left-hand
panel, and with that from \citet{gofford} in the right-hand panel. In
each panel, the solid-line histogram is that from Figure
\ref{fig:msigma} above, obtained from equation (\ref{eq:sis_result})
assuming that the black holes in the \citet{gultekin09} galaxies were
just able to drive energy-conserving supershells to the escape speeds
of their dark-matter haloes. The dashed histograms represent the AGN
data.

The most striking aspect of Figure \ref{fig:vwind_hist} is the
basic agreement, to within factors of a few at worst, in the typical
$\vw/c$ of these different samples of galaxies: our median
$\vw/c=0.035$ for the normal early-type galaxies, versus a median
$\vw/c=0.1$ for the AGN outflows of \cite{tombesi11} and a median
$\vw/c=0.056$ for the AGN of \cite{gofford}.
The overall ranges (i.e., the maxima) of the wind speeds are also very
similar. These facts are remarkable as much for the simplicity
of the model we have used to estimate $\vw/c$ in the normal galaxies,
as for the complete disconnect between the \citeauthor{gultekin09}
galaxy sample and the \citeauthor{tombesi11} or \citeauthor{gofford}
AGN samples.

To be sure, the distributions as they stand in Figure
\ref{fig:vwind_hist} are not identical. Kolmogorov-Smirnov (KS) tests
return a formal probability of
only $P_{\rm KS}\simeq 0.3\%$ that our distribution of $\vw/c$ for the
\citeauthor{gultekin09} galaxies is drawn from the same parent
distribution as the \citeauthor{tombesi11} sample, and
$P_{\rm KS}\simeq 25\%$ for equality between our $\vw/c$ values and
the \citeauthor{gofford} sample. The main reason for this appears
to be the relatively small numbers, in the present sample, of normal
galaxies with inferred $\vw/c\ga 0.1$---or, conversely, a dearth of AGN
(in the \citeauthor{tombesi11} sample especially) with slower
$\vw/c\la 0.1$.

Whatever shortcomings our very simple analysis
might have, it requires that normal galaxies with
``underweight'' black holes falling significantly below the mean
$\mbh$--$\sigma$ relation have higher-than-average $\vw/c$.
If several such galaxies were to be added to the
\citeauthor{gultekin09} sample, they could fill out the high-velocity
tail of our model $\vw/c$ distribution. 
As for the AGN, it is not clear how selection effects, observational
biases or limitations due to instrumentation may have either affected 
the measurement of relatively slow outflows, or perhaps even
prevented their inclusion in studies designed to focus on
``ultra-fast'' systems.
It is also worth noting that the probability that the $\vw/c$
measurements of \citeauthor{tombesi11} and \citeauthor{gofford} are
drawn from the same parent distribution is a formally inconclusive
$P_{\rm KS} \simeq 28\%$ ---the same as in the comparison
between the \citeauthor{gofford} distribution of $\vw$ for their AGN
and ours for the normal galaxies. As such, it is not clear that any
of the data suffice yet to allow a robust comparison at a very
detailed level between distributions of observed SMBH wind 
speeds and those inferred from any model. This makes it
even more noteworthy that the median of the $\vw$ distribution we
have obtained in this paper lies within a factor $\approx\!1.5$--3
of the median $\vw$ of two different observed distributions.

Ultimately, our results are encouraging for the general idea that
there is a parallel between the strong nuclear outflows found in
local AGN and the kind of black hole feedback that is routinely
assumed to have been a key part of galaxy formation and the
establishment of the $\mbh$--$\sigma$ relation. They also lend
support to the relevance of energy-driven feedback specifically,
and to the simple sort of modelling that we have applied to assess its
role quantitatively.

\section[]{Summary}
\label{sec:summary}

We have looked at the behaviour of energy-conserving supershells of
swept-up ambient gas driven into isothermal protogalaxies by black
hole winds. At large radii, such shells tend to a constant coasting 
speed, $v_\infty$, that depends on the black hole mass, 
$\mbh$, the black hole wind speed, $\vw$, and the velocity dispersion
of the halo, $\sigma_0$. For a shell to coast at the escape speed of a
truncated isothermal halo (i.e., $v_\infty = 2 \sigma_0$) requires
$\mbh \vw \propto \sigma_0^5$ as in equations (\ref{eq:mvwind_gamma})
and (\ref{eq:sis_result}).

We applied this escape condition for energy-conserving feedback to
the observed $\mbh$--$\sigma$ relation for the sample of quiescent
early-type galaxies and bulges of \cite{gultekin09}.  
We used equation (\ref{eq:sis_result}) to infer the
black hole wind speed that each galaxy would have had during an active
phase if our simple model is to account for the measured value of 
$\mbh$ in the galaxy, given its observed $\sigma$.  In this approach,
scatter in the 
observed $\mbh$--$\sigma$ relation directly reflects a distribution of
wind speeds from the SMBHs in the protogalaxies.  We compared the
distribution of wind velocities we obtained for the normal galaxies in
\citeauthor{gultekin09} to the observed distributions of outflow
velocities in two different samples of local AGN
\citep{tombesi11,gofford}. The distributions are strikingly
similar. Most notably, the median of our inferred wind
velocities, $\vw = 0.035c$, is within a factor $\approx\!1.5$--3 of
the median of the observed distribution of wind speeds of both
\citeauthor{tombesi11} ($\vw=0.1c$) and \citeauthor{gofford}
($\vw = 0.056c$).

\section*{Acknowledgments}

We thank J.~Reeves and J.~Gofford for helpful discussions.
We also thank the anonymous referee for useful comments.
RCM has benefited from an STFC studentship.

\label{lastpage}
\end{document}